\documentclass{article}
\usepackage{spconf,amsmath,amssymb,graphicx}
\usepackage{subcaption}
\usepackage{hyperref}

\title{Ultrafast High-Flux Single-Photon LiDAR Simulator via Neural Mapping}
\name{Weijian Zhang, Hashan K. Weerasooriya, Stanley Chan
\thanks{
The work is supported, in part, by the DARPA / SRC CogniSense JUMP 2.0 Center, NSF IIS-2133032, and NSF ECCS-2030570.
}
}
\address{School of Electrical and Computer Engineering, Purdue University, West Lafayette, USA}

\begin{document}
\topmargin=0mm
\maketitle
\begin{abstract}
Efficient simulation of photon registrations in single-photon LiDAR (SPL) is essential for applications such as depth estimation under high-flux conditions, where hardware dead time significantly distorts photon measurements. However, the conventional wisdom is computationally intensive due to their inherently sequential, photon-by-photon processing. In this paper, we propose a learning-based framework that accelerates the simulation process by modeling the photon count and directly predicting the photon registration probability density function (PDF) using an autoencoder (AE). Our method achieves high accuracy in estimating both the total number of registered photons and their temporal distribution, while substantially reducing simulation time. Extensive experiments validate the effectiveness and efficiency of our approach, highlighting its potential to enable fast and accurate SPL simulations for data-intensive imaging tasks in the high-flux regime.
\end{abstract}
\section{Introduction}
\label{sec:intro}
Single-photon LiDAR (SPL) is an emerging active imaging technique~\cite{massa_laser_1997, Rapp_2020_SPM}, particularly suitable for accurate long-range and low-light tasks because of its high sensitivity and temporal resolution empowered by advanced single-photon avalanche diode (SPAD) sensors and time-correlated single-photon counting (TCSPC) modules~\cite{Li_2021_200km_imaging}. After illuminating the scene with a periodic laser pulse train of a known shape, the SPAD and TCSPC are used to detect and register arriving photons, respectively~\cite{rochas_2003_spad, Charbon_2013_spad-based, fossum_2016_quanta, Becker_2005_TCSPC}. The depth and reflectivity of an object can be recovered from the collected photon timestamps simultaneously~\cite{kavinga_2025_joint}.

Fast and accurate simulation of photon timestamps plays a critical role in SPL for three key reasons. First, it provides a cost-effective means to study and validate the statistical properties of photons under various random processes, guiding algorithm design. Second, it allows for performance evaluation of proposed methods in a controlled environment, minimizing risks and expenses before real-world implementation. Finally, simulation is an efficient way to generate abundant training data, enabling the application of powerful deep learning techniques to SPL-related problems.

However, the challenge in rapidly and accurately simulating photons arises from the dead time in the SPAD and TCSPC, during which incoming photons cannot be detected or recorded. Without the dead time, the number of photons follows a Poisson distribution and their timestamps are independent and identically distributed (i.i.d.) ordered statistics according to the photon arrival flux function~\cite{Bar-David_1969, Snyder_1991_book}, which is determined by system and environmental parameters such as the laser repetition period, target delay, and signal strength. These factors combine in a closed-form equation that fully describes the timestamp distribution. However, when dead time is non-negligible, the registration process is disrupted by random photon losses. As a result, the photon count no longer follows a Poisson distribution, and the timestamps become dependent, governed by a complex steady-state probability density function (PDF)~\cite{Rapp_2019_Dead, rapp_2021_high}. While this PDF still depends on the same underlying parameters, the closed-form connection is lost, and the mapping becomes highly nonlinear — small changes in parameters can lead to dramatically different PDF shapes.

The conventional wisdom generates photon arrivals first and culls out those within dead time durations~\cite{Rapp_2019_Dead, weijian_2024_GUMM}. There is no need to model the photon registration number and PDF, but it is slow to examine the photons one by one. Though there exist algorithms that can correctly predict the photon registration PDFs from the parameters~\cite{Rapp_2019_Dead, rapp_2021_high}, the computation of the mapping might be time-intensive. Our goal in this paper is to model the photon count and learn the intricate mapping from the system and environmental parameters to the PDFs using a neural network, facilitating fast inference. We assume a nonparalyzable SPAD followed by a modern TCSPC module and only one dead time source as in~\cite{Rapp_2019_Dead,weijian_2024_GUMM}.

\subsection{Related Work}
\label{ssec:related_work}
\textbf{Low-flux regime.}
A typical solution is to apply the $5\%$ rule and operate the SPL system in the low-flux regime~\cite{Yu_2000_counting, Phillips_1985_5rule}. In this regime, the photon registration is equivalent to photon arrival because the probability of new photons hitting the sensor during its dead time is negligible. Therefore, techniques derived for the photon arrival process are applicable. The downside of limiting the flux to such a low level is that the data acquisition takes longer to collect enough photons to counteract the shot noise for the downstream tasks.

\noindent\textbf{Pile-up modeling.}
Under high-flux conditions, synchronous SPL systems with classic TCSPC register at most one photon per laser cycle and remain inactive until the next cycle. This leads to a bias toward early arrivals, known as the pile-up effect. Since registrations occur independently across cycles, several studies~\cite{Pediredla_2018_pileup, heide_2018_sub, Gupta_2019_Flooded} have modeled this process using a Poisson multinomial distribution, enabling efficient simulation. However, such systems suffer from slower acquisition rates compared to our asynchronous systems with modern TCSPC, which resume detection immediately after dead time. In asynchronous setups, timestamps become dependent, making the modeling of dead time distortions and thus the simulation significantly more challenging.

\noindent\textbf{Conventional simulation.}
The conventional wisdom to simulate timestamps affected by dead time distortions is based on mimicking the physics of photon registrations. A photon arrival stream is generated, followed by a sequential and manual comparison between the inter-arrival time and the dead time to decide whether to record the new photon~\cite{Rapp_2019_Dead,weijian_2024_GUMM}. While this method produces statistically accurate photon counts and timestamps, thereby circumventing the need for explicit modeling, it requires manually eliminating photons, which is time-consuming and impedes fast simulation.

\noindent\textbf{Photon registration modeling.}
To speed up the conventional simulation, one promising direction is to sample timestamps similarly to the photon arrival simulation, provided that we can fast and accurately model the registration counts and their locations. While limited work has addressed photon count modeling, a few teams~\cite{Rapp_2019_Dead,rapp_2021_high,isbaner_2016_dead} have focused on estimating the photon registration PDF, among which Rapp \emph{et al.}~\cite{Rapp_2019_Dead} rigorously modeled the photon registration process as a Markov chain. However, obtaining the PDF requires computing entries of a large transition matrix and its leading eigenvector, making the method computationally heavy. In contrast, we introduce a novel approach that models the photon count in asynchronous systems and leverages a neural network to accelerate the timestamp PDF prediction.

\subsection{Contributions: Neural Mapping}
\label{ssec:contribution}
\begin{figure}[t] 
    \centering
    \includegraphics[width=0.48\textwidth]{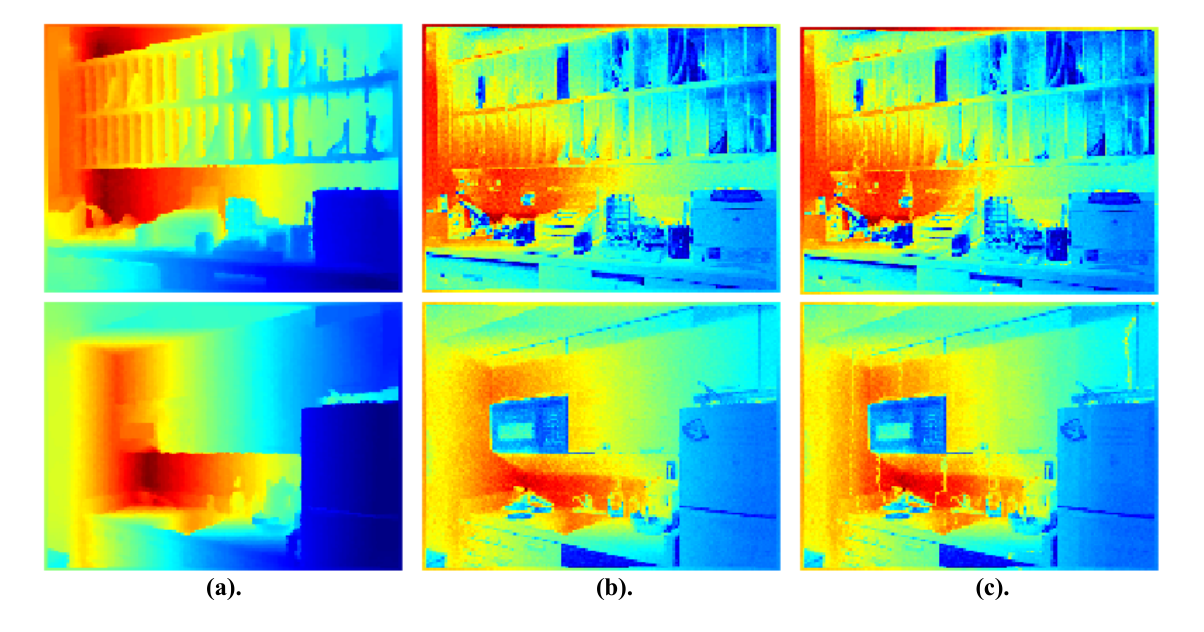} 
    \caption{Depth estimation from timestamps generated by different simulation tools: (a) Ground truth depth maps. (b) Depth estimation with the conventional simulator. (c) Depth estimation with our proposed simulator. The estimated results are visually the same for the two simulators, proving that our simulated timestamps are statistically accurate. However, it takes the conventional simulator $35.2$ minutes and ours $8.64$ seconds for a $120\times180$ image. We utilize a naive sample mean estimator for the depth recovery.}
    \label{fig:real}
\end{figure}

In this paper, we build a learning-based framework for efficient and accurate photon registration simulation in high flux. Our contributions are summarized below.
\begin{itemize}
    \item We propose a new method to model the number of photon registrations. This informs our simulator of how many photons to generate.
    \item Given system parameters, we learn a neural mapping from varying environmental parameters to corresponding distorted photon registration PDFs, following which our simulator generates samples.
\end{itemize}
Fig.~\ref{fig:real} illustrates the accuracy and speed of our simulator. The scenes are selected from NYU Depth Dataset V2~\cite{Silberman_ECCV12_dataset}.

\section{SPL Photon Registration}
\label{sec:spl_photon_registration}
In a photon timestamp simulator, the most important ingredients are the number of photons to generate and their PDF. In this section, we start from the photon arrival process and examine factors affecting photon loss and PDF distortions.
\subsection{Photon Arrival from Target}
For each pixel $(i, j)$, we assume that there is a single bounce from the object. Following~\cite{Shin_2015_3D,rapp_2017_unmixing,Chan_2024_CVPR,kao_2024_detection}, we model the per-pixel and per-cycle photon arrival as an inhomogeneous Poisson process with the flux function
\begin{equation}
    \lambda_{i,j}(t) = \alpha_{i,j} \cdot s(t - \tau_{i,j}) + \lambda_b,
    \label{eq: arrival flux}
\end{equation}
where $\tau_{i,j}$ is the pulse delay indicating the depth, $\alpha_{i,j}$ is the object reflectivity, $\lambda_b$ is the constant ambient light level across all pixels, and $s(t)$ denotes the Gaussian-shaped transmitted laser pulse, i.e. $s(t - \tau) = \calE \cdot \calN(t; \tau, \sigma_t^2)$ where $\calE$ is the laser energy and $\sigma_t$ is the half pulse width.

For simplicity, we remove the subscript $(i,j)$ and make the following assumptions. The repetition period $t_r$ is longer than the delay from the farthest target. The pulse width is negligible compared to the repetition period such that $\int_0^{t_r} s(t - \tau) \, dt \approx \calE$. The sensor's quantum efficiency $\eta = 1$ and the dark current $\lambda_d = 0$.

Within one cycle, the energy carried by $\lambda(t)$ is
\begin{equation}
    Q \bydef \int_0^{t_r} \lambda(t) \, dt = \alpha \calE + t_r\lambda_b = S + B,
\end{equation}
where we define the signal level $S \bydef \alpha \calE$, the noise level $B \bydef t_r\lambda_b$. We define the signal-to-background ratio (SBR) as $\text{SBR} \bydef S / B$.

\subsection{Photon Arrival Simulator}
\label{ssec:arrival simulator}
We can follow the two steps below to simulate photon arrival timestamps during $N$ acquisition cycles as in \cite{Chan_2024_CVPR}.
\begin{itemize}
    \item Step 1: Decide on the number of photon arrivals $M_a$ to be generated. $M_a$ is a Poisson random variable whose parameter is the total energy $N Q$. Thus, we simply draw a sample based on
    \begin{equation}
        \label{eq: count sampling}
        M_a \sim \texttt{Poisson}(NQ).
    \end{equation}
    \item Step 2: Generate $M_a$ timestamps that statistically match Eq.~\eqref{eq: arrival flux}. In the photon arrival process, photons within the same cycle and even from different cycles are i.i.d. according to the following PDF
    \begin{equation}
        f_a(t) = \frac{\lambda(t)}{Q} = \frac{\alpha}{S+B} \cdot s(t - \tau) + \frac{\lambda_b}{S+B},
        \label{eq: arrival pdf}
    \end{equation}
    if we apply the relative timestamp representation, i.e $t \in [0, t_r)$. Then, the $M_a$ timestamps can be immediately sampled from Eq.~\eqref{eq: arrival pdf} using the inverse transform method since we have the knowledge of the PDF.
\end{itemize}

\subsection{Photon Registration at Sensor}
Photon arrival simulation is efficient because the mappings to $M_a$ and $f_a(t)$ are analytically defined by Eqs.~\eqref{eq: count sampling} and~\eqref{eq: arrival pdf}. However, dead time incurs random photon loss, making the mappings to the photon registration count $M_r$ and PDF $f_r(t)$ highly nonlinear and difficult to simulate. We utilize the conventional simulation scheme to study the functional connections from system/environmental parameters to $M_r$ and $f_r(t)$.

\subsubsection{Number of photon registrations}
\label{sssec:reg count}
Rather than the system parameters $\{t_r, t_d, \sigma_t, N\}$, which are determined by the hardware, we are more interested in understanding how the scene-dependent environmental parameters $\{\tau, S, B\}$ influence the transition from $M_a$ to $M_r$.

\begin{figure}[t]
    \centering
    \begin{subfigure}[b]{0.49\columnwidth}
        \centering
        \raisebox{2.3pt}{\includegraphics[width=\linewidth]{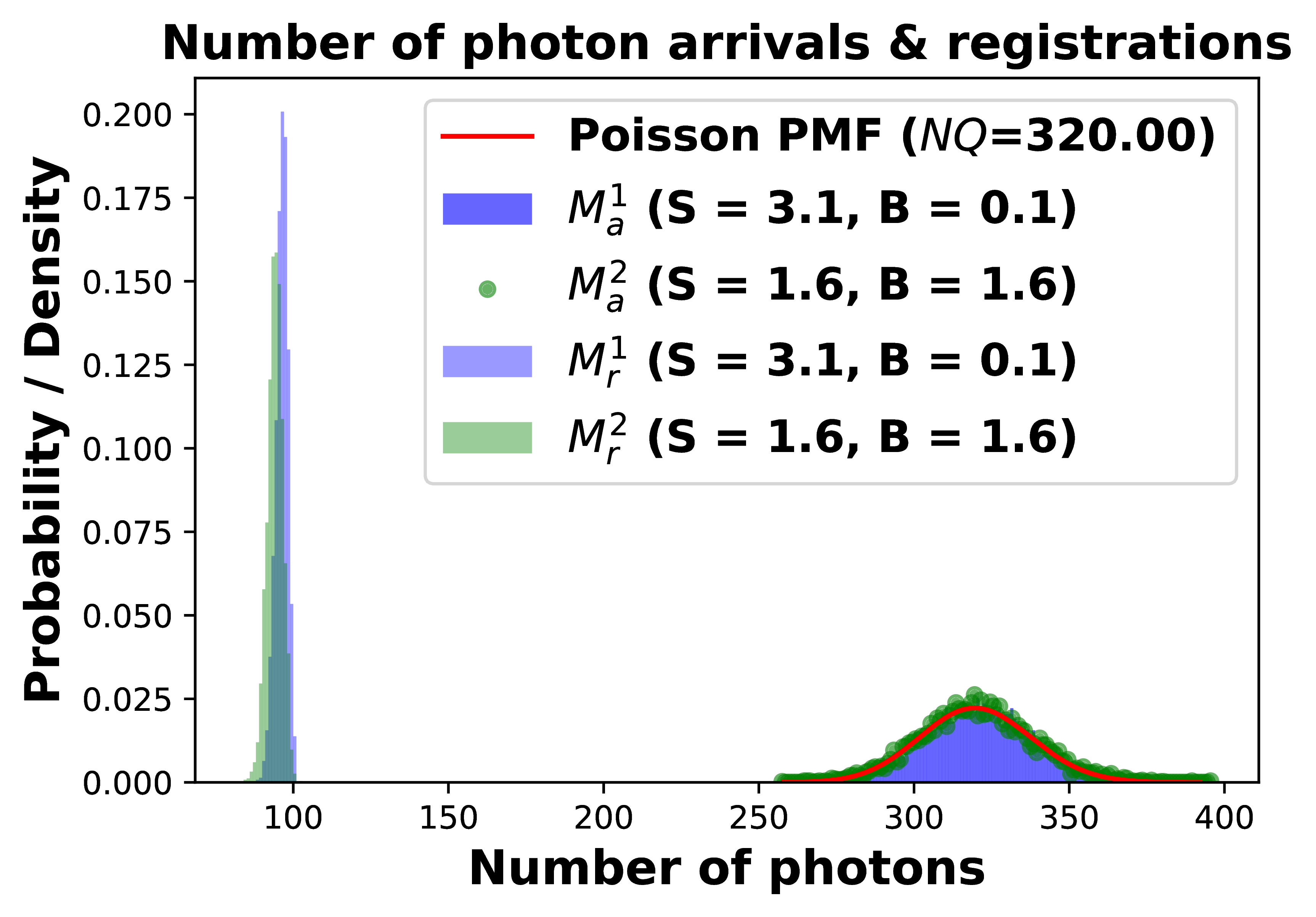}}
        \caption{Lower total energy ($NQ$).}
        \label{fig:2a}
    \end{subfigure}
    \hfill
    \begin{subfigure}[b]{0.49\columnwidth}
        \centering
        \includegraphics[width=\linewidth]{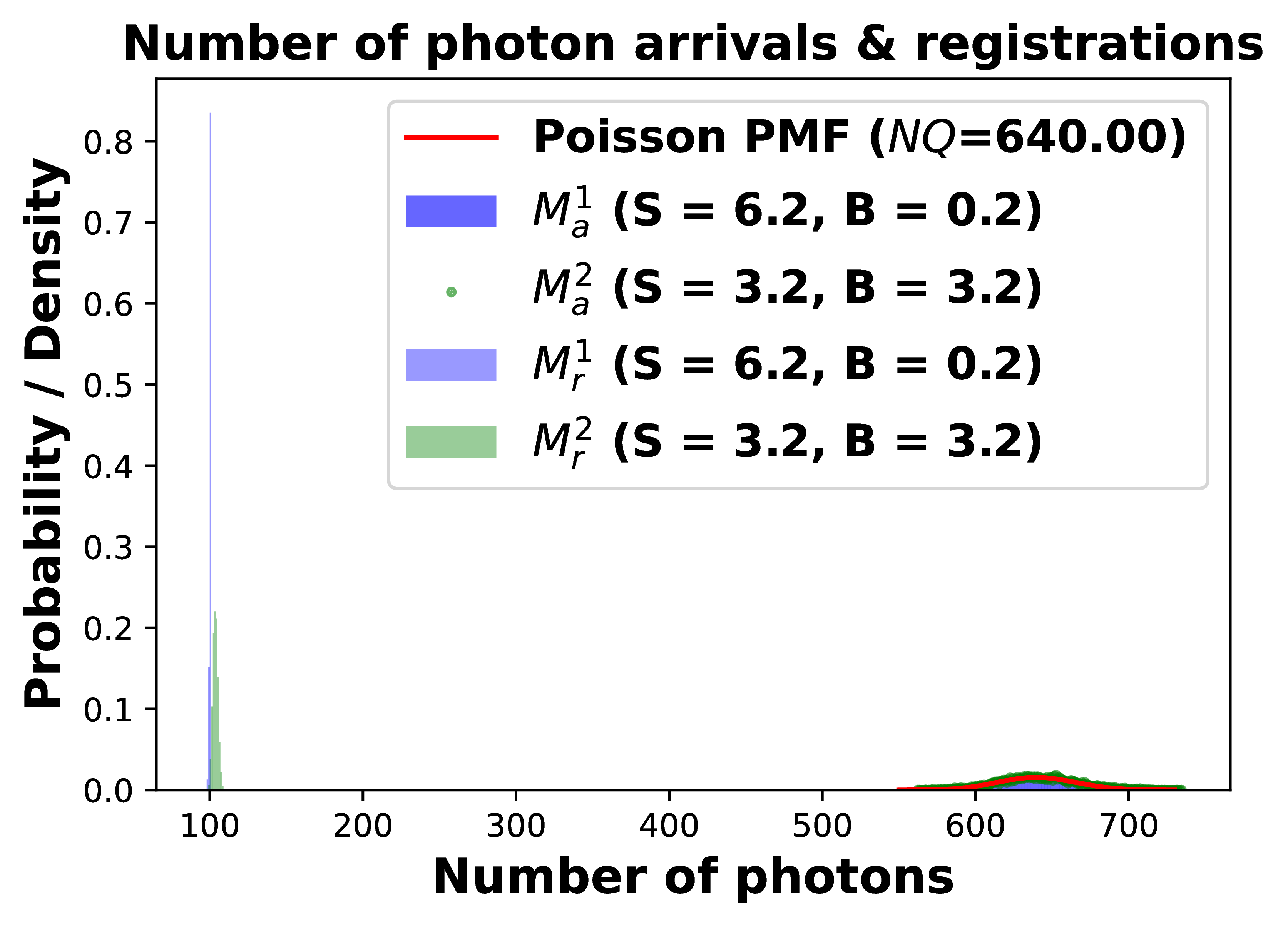}
        \caption{Higher total energy ($NQ$).}
        \label{fig:2b}
    \end{subfigure}
    \caption{Empirical distribution of photon arrival and registration counts under varying energy and SBRs.}
    \label{fig:simulation_num_regs}
\end{figure}

We simulate $5000$ realizations of photon arrival and the consequent registration process under different total energy levels, as shown in Fig.~\ref{fig:2a} and~\ref{fig:2b}. For each energy level, we vary the SBR and build empirical histograms of arrival counts $(M_a^1, M_a^2)$ and registration counts $(M_r^1, M_r^2)$ under different SBRs. Both subfigures illustrate that while $M_a^1$ and $M_a^2$ are statistically equivalent, $M_r^1$ and $M_r^2$ exhibit distinct distributions. We observe that $M_r$ is bell-shaped with reduced mean and variance compared to $M_a$, though the precise relationship remains unknown.

From Fig.~\ref{fig:2a} to Fig.~\ref{fig:2b}, we fix the SBR while doubling the energy. The mean and variance of $M_r$ do not scale linearly. The mean remains nearly constant, as the hardware’s ability to register photons is constrained by the fixed dead time duration, emphasizing the complexity of modeling $M_r$.

\subsubsection{Photon registration PDF}
\begin{figure}[t]
    \centering
    \begin{subfigure}[b]{0.95\columnwidth}
        \centering
        \includegraphics[width=\linewidth]{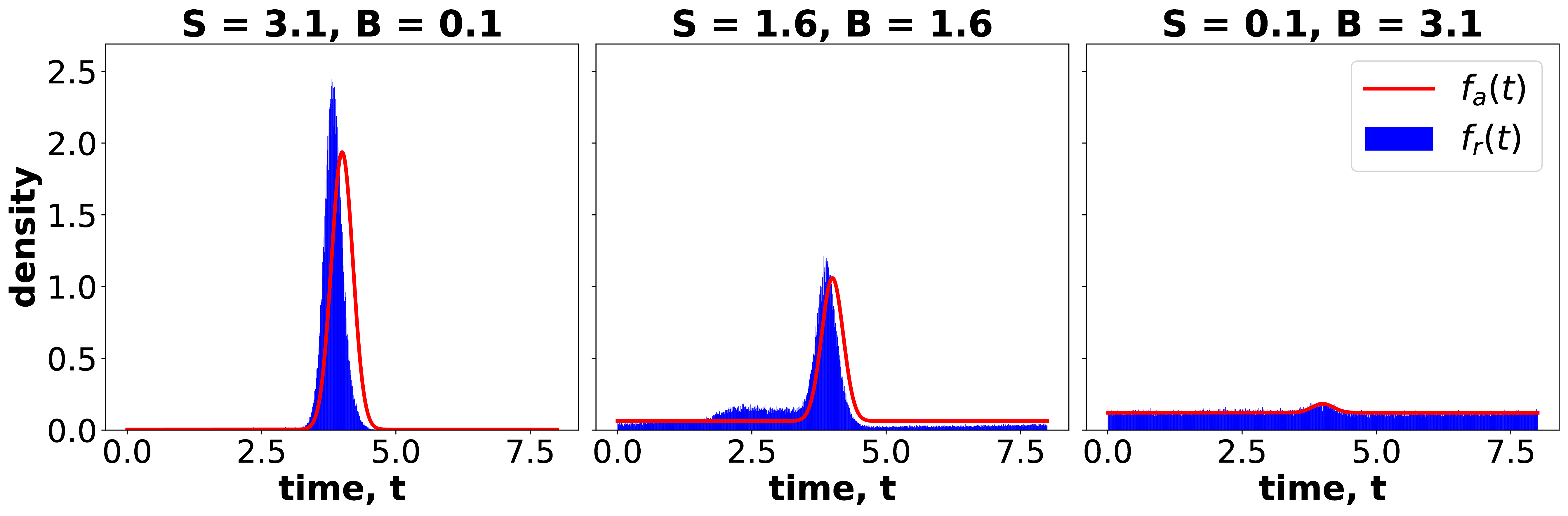}
        \caption{Fixed energy, different SBRs.}
        \label{fig:reg_pdf_sim_sbr}
    \end{subfigure}
    
    \begin{subfigure}[b]{0.95\columnwidth}
        \centering
        \includegraphics[width=\linewidth]{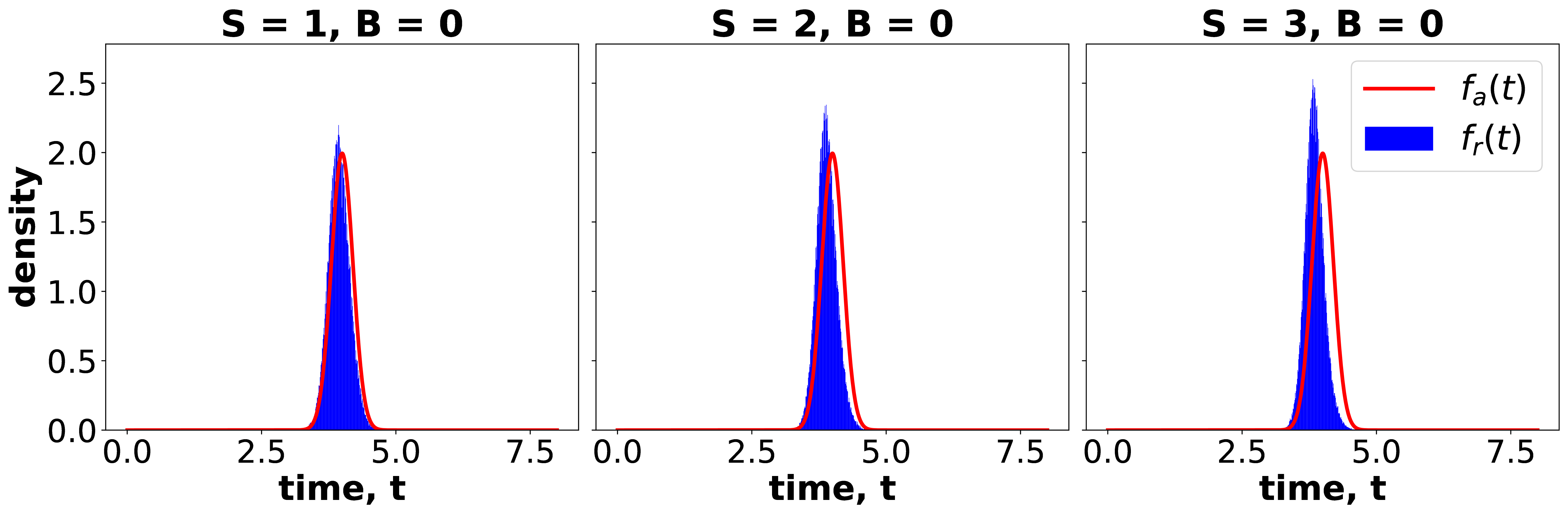}
        \caption{Fixed SBR, different energy.}
        \label{fig:reg_pdf_sim_energy}
    \end{subfigure}
    \caption{Effects of signal and background levels on the registration PDF $f_r(t)$.}
    \label{fig:reg_pdf_sim}
\end{figure}

Similarly, we simulate photon registrations for $20$ realizations and average the empirical histograms to approximate the ground truth PDF $f_r(t)$. We investigate the effects of various $S$ and $B$ on $f_r(t)$ in Fig.~\ref{fig:reg_pdf_sim}.

\textbf{Fixed energy, different SBRs.} As shown in Fig.~\ref{fig:reg_pdf_sim_sbr}, SBR affects the weights of three components: a main peak, a small bump, and a noise floor. As the SBR decreases, the main peak shrinks, the noise floor grows, and a small bump emerges because noise-incurred photons accumulate roughly $(t_r - t_d)$ ahead of the main peak, resembling pile-up effects.

\textbf{Fixed SBR, different energy.} Fig.~\ref{fig:reg_pdf_sim_energy} indicates that increasing total energy, with fixed SBR (infinity as a special case), leads to greater distortion toward earlier timestamps. This occurs because the system’s capability of registering photons is limited by the fixed dead time duration. As energy increases, more photons arriving after $\tau$ are missed, resulting in a pronounced leftward shift of the main peak.

In conclusion, $S$ and $B$ control the relative proportions of the main peak, noise bump, and noise floor, as well as the degree of distortion. Meanwhile, $\tau$ shifts the entire histogram horizontally. Therefore, deriving the registration PDF directly from these three scene parameters is highly nontrivial.

\section{Method}
\label{sec:method}
In this section, we formulate the problems and propose our solutions to the challenging mapping tasks of the photon registration number $M_r$ and PDF $f_r(t)$, respectively.

\subsection{Problem Statement}
From the previous section, $M_r$ and $f_r(t)$ both highly depend on a set of parameters $\boldsymbol{\theta} = \{t_r, t_d, \sigma_t, N, \tau, S, B\}$. We partition $\boldsymbol{\theta}$ into two groups: system parameters $\boldsymbol{\theta_s} = \{t_r, t_d, \sigma_t, N\}$ that rely on the laser and sensor specifications and environmental parameters $\boldsymbol{\theta_e} = \{\tau, S, B\}$ that come from the scene.

For most SPL applications, the system is calibrated and $\boldsymbol{\theta_s}$ is known. Therefore, the goal is to model and estimate the mapping from $\boldsymbol{\theta_e}$ to the corresponding photon count $M_r$ and PDF $f_r(t)$, given $\boldsymbol{\theta_s}$.

\subsection{Modeling of Photon Registration Count}
$M_r$ is the result of subtracting photon loss from the arrival photon count $M_a$. For each photon registered at $t_k$, the number of missing photons in the subsequent dead time is proportional to the energy under the flux function $\lambda(t)$ from $t_k$ to $t_k + t_d$. If $t_k + t_d > t_r$, we apply a periodic padding to build an extended flux function $\Tilde{\lambda}(t)$.

\begin{figure}[t]
    \centering
    \begin{subfigure}[b]{0.95\columnwidth}
        \centering
        \includegraphics[width=\linewidth]{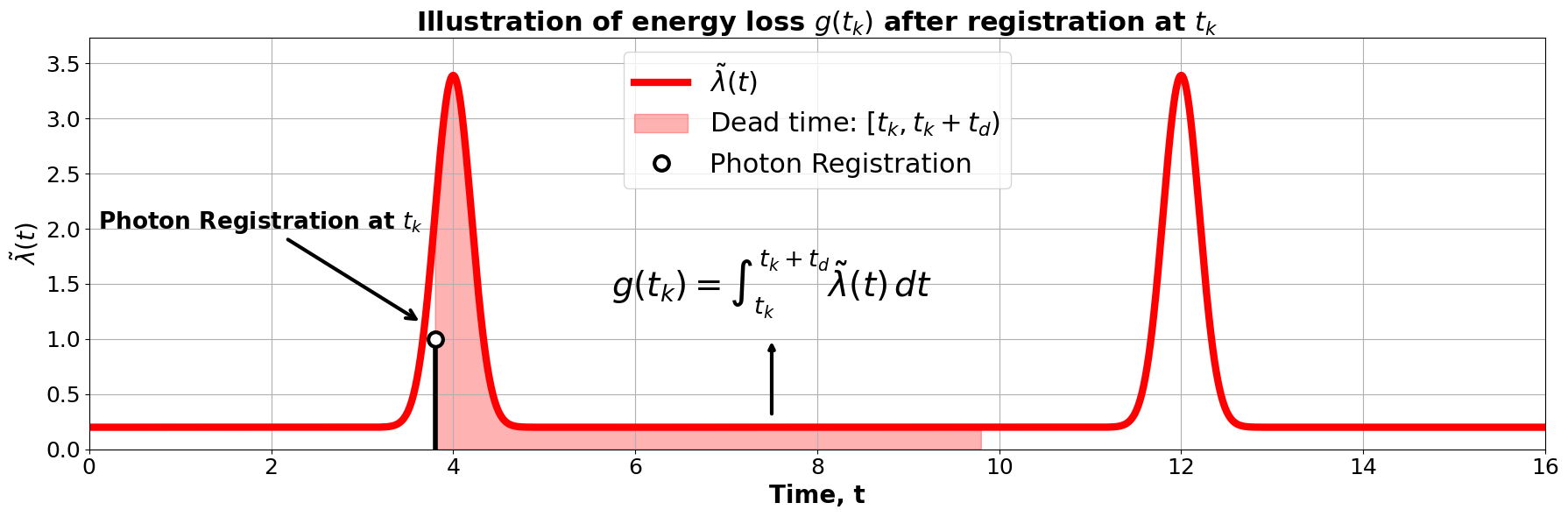}
        \caption{Energy loss per registration modeled by integration over the extended flux function $\Tilde{\lambda}(t)$.}
        \label{fig:reg_num_method1}
    \end{subfigure}
    
    \begin{subfigure}[b]{0.95\columnwidth}
        \centering
        \includegraphics[width=\linewidth]{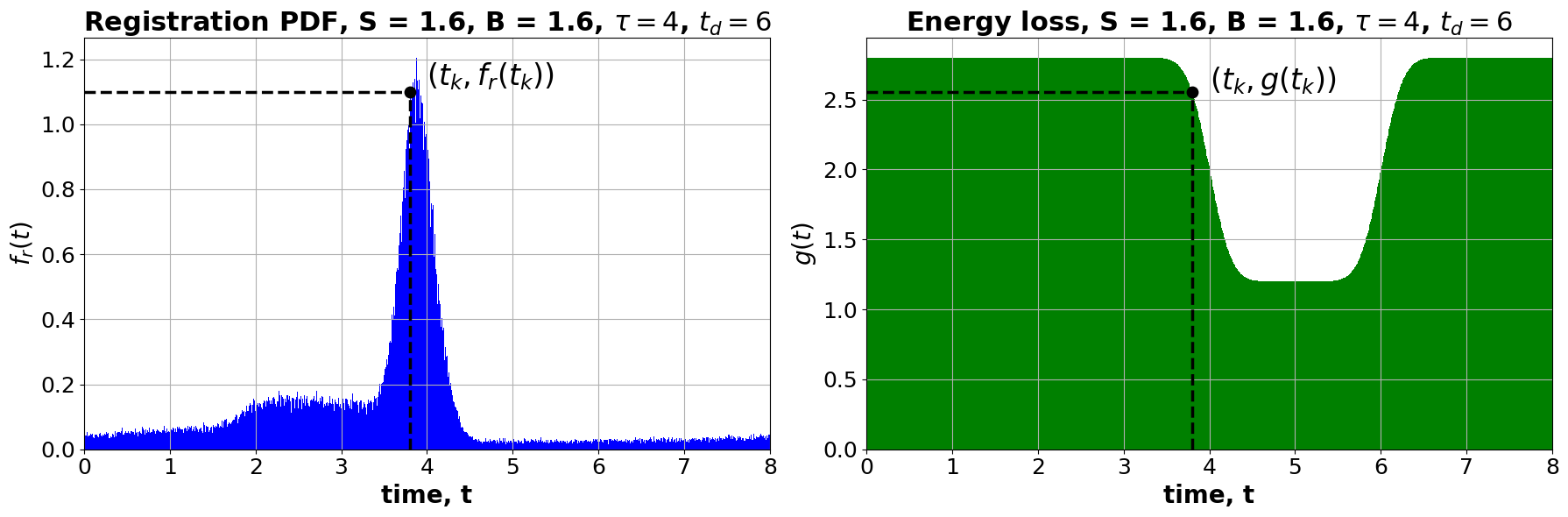}
        \caption{Expected photon loss computed as the inner product. In this case, $\E[g(t_k)] = \langle f_r,g  \rangle = 2.43$.}
        \label{fig:reg_num_method2}
    \end{subfigure}
    \caption{Visualization of energy loss modeling for estimating photon registration count.}
    \label{fig:reg_num_method}
\end{figure}

We define a function $g(\cdot)$ that maps from a timestamp to the energy loss starting at it, then $g(t_k) = \int_{t_k}^{t_k+t_d} \Tilde{\lambda}(t) \ dt$, as illustrated in Fig.~\ref{fig:reg_num_method1}. Since $t_k$ follows the stationary photon registration PDF $f_r(t)$, the expected number of photon loss per new registration is
\begin{equation}
    \E[g(t_k)] = \int_0^{t_r} g(t_k) f_r(t_k) \ dt_k = \langle f_r,g  \rangle,
\end{equation}
which is a dot product between $f_r(\cdot)$ and $g(\cdot)$, as in Fig.~\ref{fig:reg_num_method2}.

Accounting for the photon loss, we construct an estimator for $M_r$ as $M_a / \left(1 + \E[g(t_k)]\right)$. When $NQ$ is large, $M_a \sim \texttt{Poisson}(NQ) \approx \calN(NQ, NQ)$. Therefore,
\begin{align*}
    \frac{M_a}{1 + \E[g(t_k)]}
    & \sim \calN \left(\frac{NQ}{1 + \E[g(t_k)]}, \frac{NQ}{(1 + \E[g(t_k)])^2} \right) \\
    & \sim \calN \left(R, \frac{R}{1 + \E[g(t_k)]} \right),
\end{align*}
where $R \bydef NQ / \left(1 + \E[g(t_k)]\right)$. 

This is a valid Gaussian estimator that is consistent with the bell-shaped observations in Section~\ref{sssec:reg count}. However, the empirical width is narrower than $\sqrt{\frac{R}{1 + \E[g(t_k)]}}$. Therefore, we refine the estimator as
\begin{equation}
    \label{eq: count estimator}
    \widehat{M_r} \sim \calN \left(R, \frac{R}{(1 + \E[g(t_k)])^2} \right).
\end{equation}
The only unknown in Eq.~\ref{eq: count estimator} is $f_r(t)$, which is solved in the next subsection.

\subsection{Neural Mapping of Photon Registration PDF}
\label{ssec:pdf_prediction}
The mapping from environmental parameters $\boldsymbol{\theta_e}$ to the photon registration PDF $f_r(t)$ is highly nonlinear and difficult to model analytically. Our goal is to approximate this mapping efficiently using a neural network.

We explored various input-output representations. One option is to directly discretize $f_r(t)$ into a high-dimensional vector, which aligns with the need for fine temporal resolution in timestamp sampling. However, $\boldsymbol{\theta_e}$ is a low-dimensional vector, and this dimensionality mismatch impedes effective learning. Alternatively, we considered representing $f_r(t)$ using a Gaussian-uniform mixture model (GUMM) to reduce output dimensionality~\cite{weijian_2024_GUMM}. Yet, the ambiguity of GUMM parameters led to poor convergence during training.

\begin{figure}[t]
    \centering
    \includegraphics[width=0.48\textwidth]{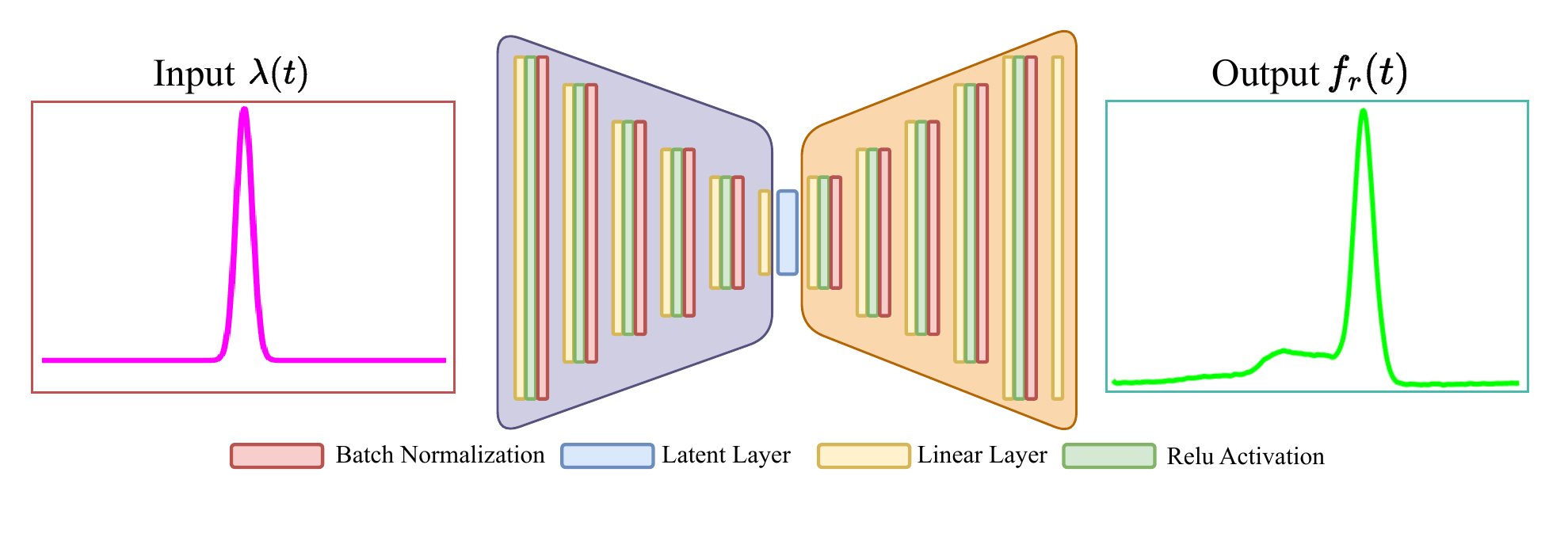} 
    \caption{Proposed AE architecture for PDF's neural mapping.}
    \label{fig:netowk_architect}
\end{figure}

To address this, we observe that $\boldsymbol{\theta_e}$ is implicitly embedded in the photon arrival flux function $\lambda(t)$, as in Eq.~\ref{eq: arrival flux}. Discretizing $\lambda(t)$ to the same dimension as $f_r(t)$ resolves the mismatch. Based on this insight, we design an autoencoder (AE) to extract features from $\lambda(t)$ and reconstruct the corresponding distorted PDF $f_r(t)$, as shown in Fig.~\ref{fig:netowk_architect}.

\section{Experimental Results}
\label{sec:results}

\subsection{Network and Training Specifications}
A simple yet effective AE architecture is proposed in Fig.~\ref{fig:netowk_architect}. The input flux function $\lambda(t)$ and the output distorted PDF $f_r(t)$ are discretized into $1024$-dimensional vectors, which can be adjusted based on the actual bin resolution of TCSPC. The encoder progressively downsamples by a factor of $2$ per layer, while the decoder symmetrically upsamples, with a latent representation of size 16 at the bottleneck.

We generate a dataset of $11,000$ input-output pairs for training and testing, with a $4:1$ split. Physically meaningful environmental parameters $\boldsymbol{\theta_e}$ are uniformly sampled from $S \sim \mathcal{U}(0,3)$, $B \sim \mathcal{U}(0,3)$ and $\tau \sim \mathcal{U}(2,6)$ to calculate the flux function as defined in Eq.~\ref{eq: arrival flux}. For each sample, $20$ realizations of photon registrations are simulated using the conventional wisdom, and their averaged histogram serves as the empirical ground truth PDF (label).

We adopt the mean squared error (MSE) between the AE network's prediction and the ground truth as the loss function. The network is trained using the Adam optimizer with a batch size of $128$ for $5000$ epochs on an NVIDIA A100 GPU.

\subsection{Results for the Simulated Dataset}
\begin{figure}[t]
    \centering
    \begin{subfigure}[b]{0.49\columnwidth}
        \centering
        \includegraphics[width=\linewidth]{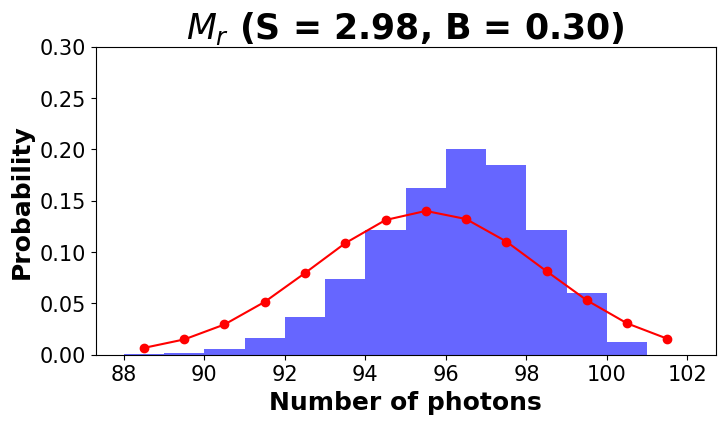}
    \end{subfigure}
    \hfill
    \begin{subfigure}[b]{0.49\columnwidth}
        \centering
        \includegraphics[width=\linewidth]{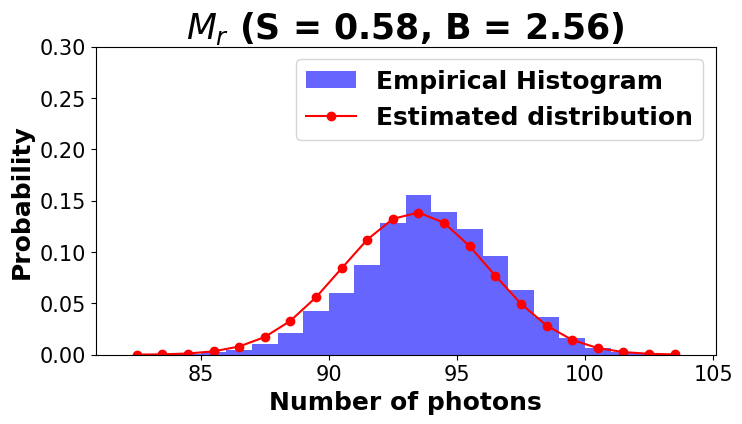}
    \end{subfigure}
    \caption{Prediction of photon registration counts.}
    \label{fig:count_results}
\end{figure}

\begin{figure}[t] 
    \centering
    \includegraphics[width=0.48\textwidth]{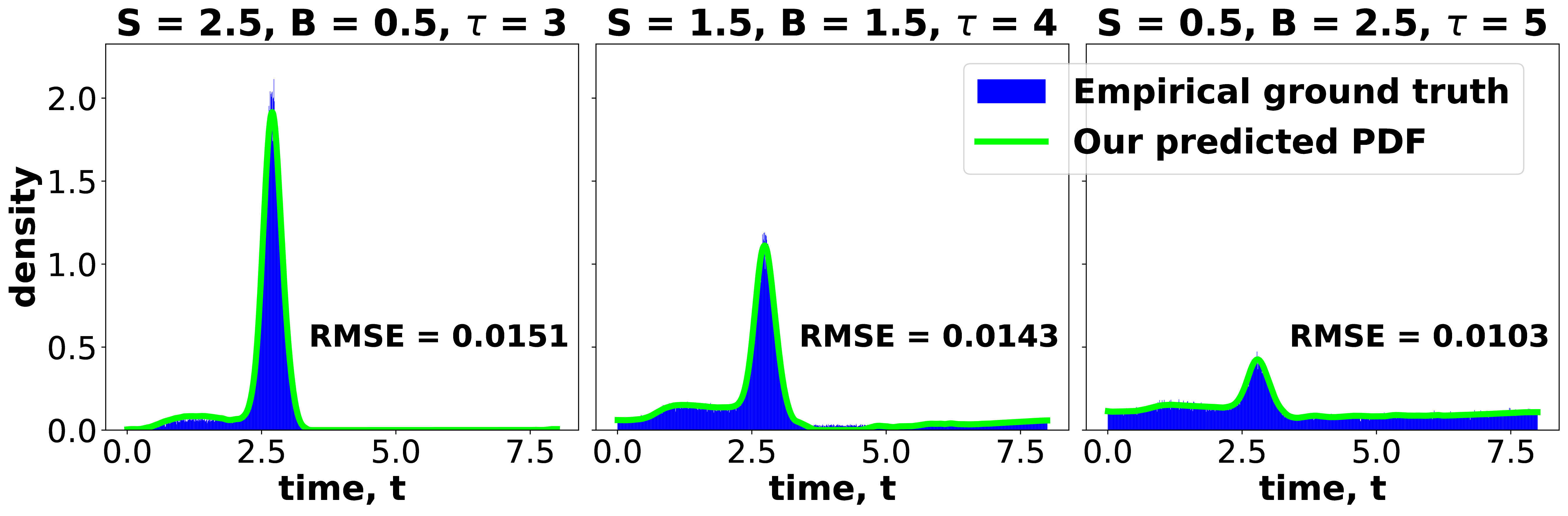} 
    \caption{Prediction of photon registration PDFs.}
    \label{fig:results}
\end{figure}

Figs.~\ref{fig:count_results} and~\ref{fig:results} illustrate the effectiveness of our method by comparing the predicted photon registration counts and PDFs with empirical ground truth, respectively, across various combinations of environmental parameters $\boldsymbol{\theta_e}$. The proposed AE accurately captures the distorted PDFs, which in turn enables precise estimation of the registration counts.

The root mean square error (RMSE) of the PDF prediction across the testing dataset is $0.017$. Notably, the handpicked $\boldsymbol{\theta_e}$ values in Fig.~\ref{fig:results} lie outside the dataset, reflecting the generalizability of the network.

\subsection{Runtime Analysis}
\begin{figure}[t] 
    \centering
    \includegraphics[width=0.48\textwidth]{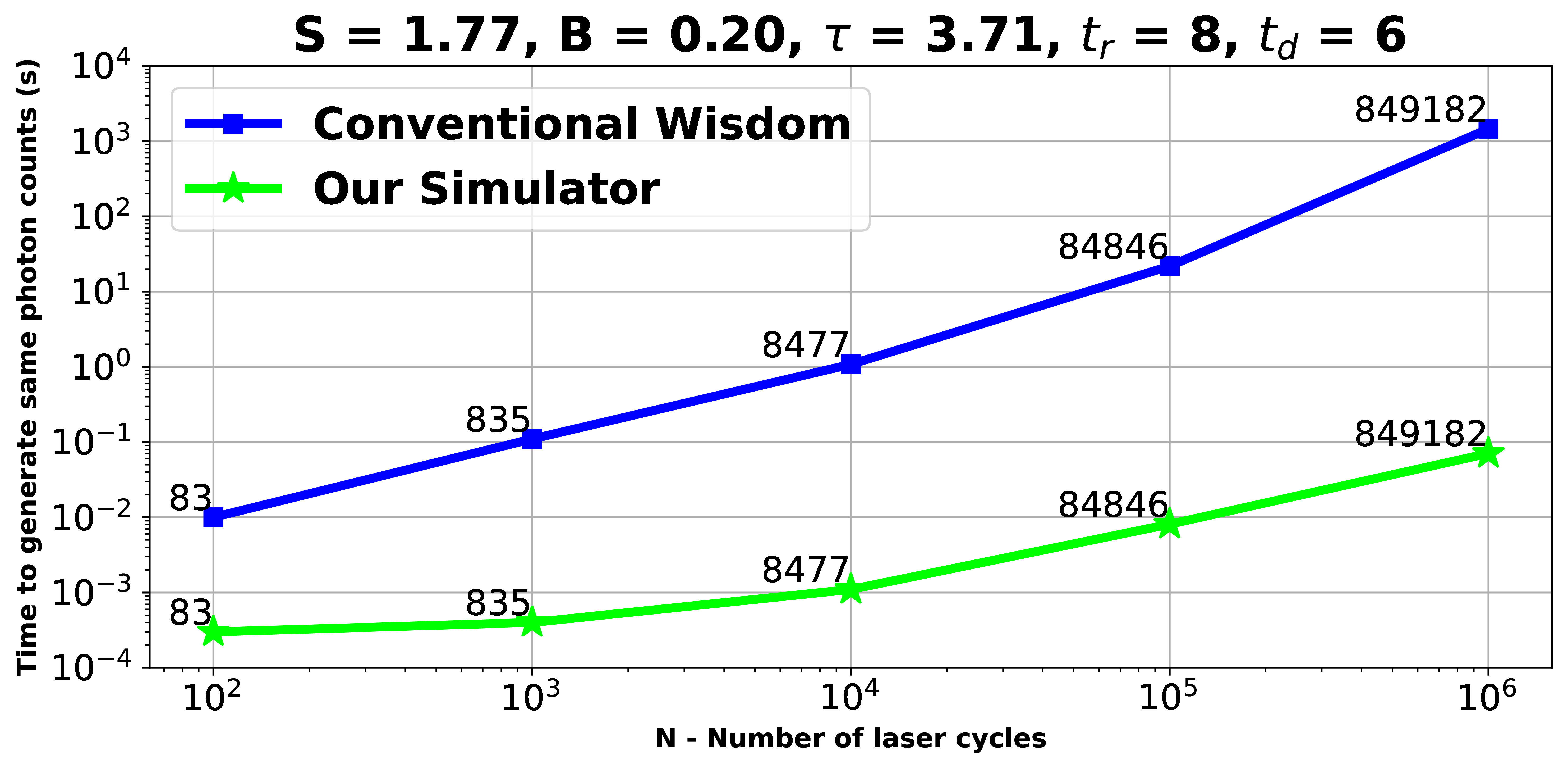} 
    \caption{Per-pixel runtime comparison between conventional and proposed simulators.}
    \label{fig:time_results}
\end{figure}

We further emphasize the efficiency of our method by measuring the time each simulator requires to generate per-pixel timestamps, as shown in Fig.~\ref{fig:time_results}. The values annotated along the curves indicate the effective number of registered photons, accounting for losses due to dead time.\footnote{Experiments were conducted in MATLAB 2023b on a system with an Intel\textsuperscript{\textregistered} Core\texttrademark{} i7-8700K CPU and 32\,GB DDR4 RAM.}

While the runtime of the conventional simulator increases rapidly with the number of laser cycles, our method maintains negligible latency, demonstrating its suitability for efficiently simulating large-scale SPL images and videos.





\section{Conclusion}
\label{sec:conclusion}
In this work, we introduce a learning-based approach to accelerate the simulation of photon registrations in high-flux single-photon LiDAR systems, where detector dead time introduces complex distortions to photon statistics. We design a new model for the registration count and learn a neural mapping from environmental parameters to the registration PDF. Our method effectively bridges the gap between conventional, computationally expensive simulations and the need for large-scale training data in deep learning applications. Experimental results demonstrate that our approach achieves high accuracy in both photon count and PDF estimations while significantly reducing computational time.


\bibliographystyle{IEEEbib}
\bibliography{refs}

\end{document}